# Charge Separation and Dissipation in Molecular Wires under a Light Radiation

Hang Xie[*1], Yu Zhang[2], Yanho Kwok[3], Wei E.I. Sha[4]

[1] Department of Physics, Chongqing University, Chongqing, China

[2] Department of Chemistry, Northwestern University, Evanston, Illinois, United States

[3] Department of Chemistry, the University of Hong Kong, Hong Kong, China

[4] College of Information Science & Electronic Engineering, Zhejiang University, Hangzhou 310027, P. R. China

Photo-induced charge separation in nanowires or molecular wires had been studied in previous experiments and simulations. Most researches deal with the carrier diffusions with the classical phenomenological models, or the static energy levels by quantum mechanics calculations. Here we give a dynamic quantum investigation on the charge separation and dissipation in molecule wires. The method is based on the time-dependent non-equilibrium Green's function theory. Polyacetylene chain and poly-phenylene are used as model systems with a tight-binding Hamiltonian and the wide band limit approximation in this study. A light pulse with the energy larger than the band gap is radiated on the system. The evolution and dissipation of the non-equilibrium carriers in the open nano systems are studied. With an external electric potentials or impurity atoms, the charge separation is observed. Our calculations show that the separation behaviors of the electron/hole wave packets are related to the Coulomb interaction, light intensity and the effective masses of electron/hole in the molecular wire.

## I. INTRODUCTION

In the development of photovoltaic devices, some nanostructures such as graphene, silicon-nanowires, and 2D materials have gained a lot of interest [1, 2]. In these nanostructures, their energy band structures could be modulated by the chemical doping, geometric cutting and the strains on them [3-6]. Thus the power convention efficiency (PCE) will also be improved by many of these factors. To increase the PCE of nanostructures, the charge separation is a very important process. It can be realized in the p-i-n typed nanowires by different atom doping or the bias potentials exerted on two regions. On the other hand, the strains in nanowires would change their band structures, which lead the spontaneous electron-hole separation in the real space[5].

Regarding the experimental aspect, recently the pump-probe technique in the ultrafast optics is used to directly analyse the transient charge separation process in silicon nanowires [6-8]. In the bending nanowires, the induced strain in them also increases the electron-hole recombination rate[6].

On the theoretical and computational aspect, the traditional drift-diffusion equation describes the carries movement under the light illumination in a large scale [7-11]. In this classical theory, some parameters (e.g. carrier recombination rate and life time) are obtained from some phenomenological models such as the Shockley-Read-Hall model [9]. In nano-scale, we have to describe these phenomena by quantum mechanics. For example, the Bethe-Saplter equation (BSE) is often employed for the exciton generation [12, 13]. Nowadays many first-principles calculations, such as the density functional theory (DFT), are employed for studying the energy band of nanowires under light illumination [13-16]. Combined with the non-equilibrium Green's function (NEGF) theory, DFT calculations are also employed for the nano-structured photovoltaics [17, 18]. For molecular wires, the charge transfer and separation are also investigated by DFT or other quantum chemistry calculations [19, 20].

Among these works, the calculation for the dynamic charge separation process is still unexplored in the nano-scale devices, based on the ballistic quantum calculation. Recently we have developed a set of hierarchical equations of motion method (HEOM) to study the time-dependent quantum transport [21-24]. This method may be also employed to study the nano structured photovoltaics. In this paper we adopt this tool to investigate the transient charge separation of nanowires directly in time and space domain. We find that the bias potential or the doping atoms in the molecular wires can induce the electron-hole separation after a laser radiates in the molecular wire. Under such radiation, the carrier dissipation and wave packet speed in different doping situations, Coulomb interaction and band structures are also investigated for these open systems. We also use our theory to give a clear description for the quantum dissipation process in these systems.

## II. MODEL AND THEORY
### A. Tight binding model and transport schemes

In this paper we use the polyacetylene (PA) and poly-(p-phenylene) (PP) or doped PP as the research systems. For PA, due to the Peierls distortion [25], the polyacetylene has the alternative single and double bonds. The trans-form (Fig. 1(a)) is more thermodynamically stable than the cis-form. In our calculation, we use the nearest neighbor tight-binding (TB) model for the carbon atoms with two hopping integral parameters ($t_1$=-2.7 eV and $t_2$=-2.0 eV) to represent the two types of bonds (Fig. 1(a)). For PP chains, we also only consider the carbon (nitrogen and boron) atoms in the nearest neighbor TB model with a single hopping integral ($t$=-2.7 eV) (Fig. 1(b) and Fig. 1(2)). For the doped PP chain, the on-site energies of the atoms are changed properly to fit for the first-principles band results.

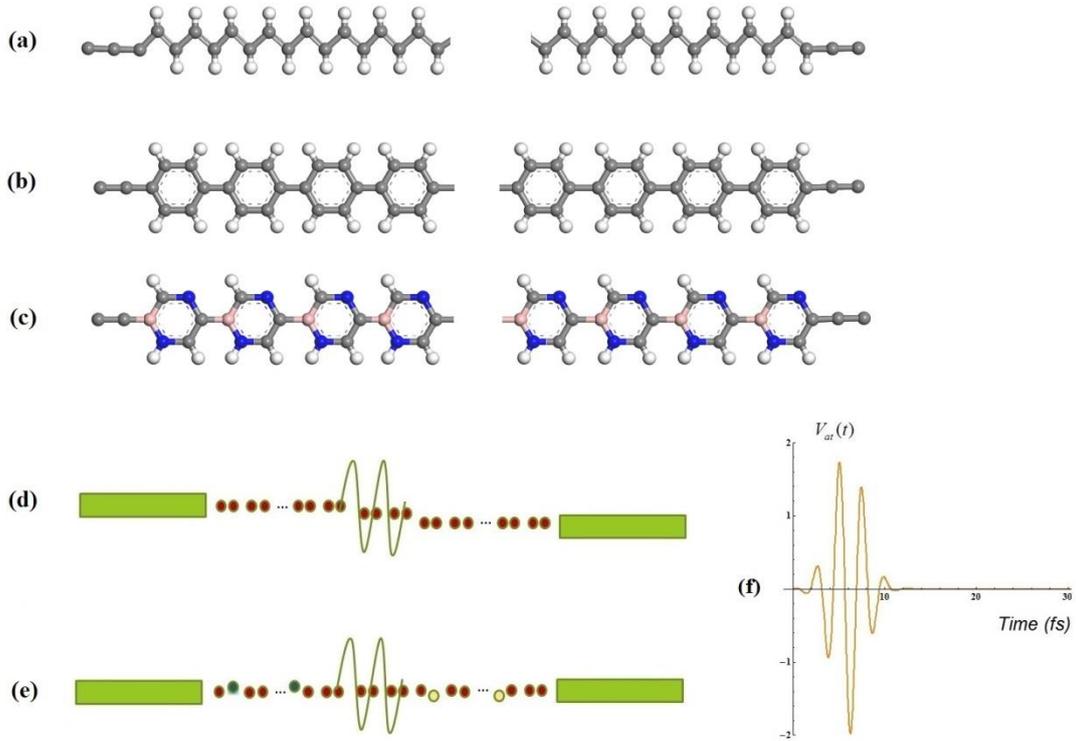

**FIG. 1.** (a) Atomic figure for the PA chain; (b) Atomic figure for the PP chain; (c) Atomic figure for PP chain doped with boron (pink) and nitrogen (blue) atoms; In (a)-(c), two carbon-atom chains sandwich the middle molecular wires for the WBL condition. (d) and (e) Schematic picture for the molecular wire system under a laser radiation with impurity atoms. The sine curves in the middle represent the laser beams radiated on the p-i-n junction of the chain system. In (d) the impurity atoms are represented by two averaged bias potentials on both sides; while in (e) the impurities are represented by the single impurity atoms with positive or negative on-site potentials. (f) The time dependence of the laser pulse.

In the transport calculation for the exciton generation and charge separation, we choose two schemes for the doping effect. In the first scheme we artificially adjust the on-site energy of each atom with the averaged potentials in the p-doped or n-doped regions (Fig.1 (d)). In the second scheme we change the on-site potentials for some atoms to represent the two types of doping (Fig. 1(e)). To investigate the dynamic charge separation, we set a laser pulse to radiate in the middle region of the PA/PP chain and then calculate the electron evolution in the open system by the method below. Due to the computational limitation, the radiation region is set to be much smaller than the actual spot size of the laser beam. However, our calculation results show that such a small radiation region is enough to observe the charge separation process in the molecular wires.

## B. HEOM -WBL theory

In this study, we employ our developed HEOM theory [21, 23] to solve the electron dynamics for the molecular wires. Due to the large size of the long molecular wire, we also utilize the wide-band-limit (WBL) approximation in the calculation. A brief introduction for this HEOM-WBL theory is given below.

In the quantum transport calculations, an open system is partitioned into three regions: the left lead ($L$), the device ($D$) and the right lead ($R$). The equation of motion for the density matrix of the device is

$$i\dot{\boldsymbol{\sigma}}(t) = [\mathbf{h}_D(t), \boldsymbol{\sigma}(t)] - \sum_{\alpha}^{N_\alpha}[\boldsymbol{\varphi}_\alpha(t) - \boldsymbol{\varphi}_\alpha^\dagger(t)], \qquad (1)$$

where $\boldsymbol{\sigma}_D(t)$ and $\mathbf{h}_D(t)$ are the single-electron density matrix and Hamiltonian of the device. $\boldsymbol{\varphi}_\alpha(t)$ is called the auxiliary density matrix, which corresponds to the dissipation effect from the leads. With the residue theory, the auxiliary density matrix can be written as [21]

$$\boldsymbol{\varphi}_\alpha(t) = i(\boldsymbol{\sigma}(t) - \frac{\mathbf{I}}{2}) \cdot \Lambda_\alpha^z + \sum_{k=1}^{N_k} \boldsymbol{\varphi}_{\alpha,k}(t), \qquad (2)$$

where $\boldsymbol{\varphi}_{\alpha,k}(t)$ is defined as the discrete auxiliary density matrix. Its detailed expression can be found in other references [21, 23]. The EOM of $\boldsymbol{\varphi}_{\alpha,k}(t)$ is derived as

$$i\dot{\boldsymbol{\varphi}}_{\alpha,k}(t) = \frac{2iR_k}{\beta}\boldsymbol{\Lambda}_\alpha^z + [\mathbf{h}_D(t) - (\Delta_\alpha(t) + z_k/\beta)\mathbf{I} - i\boldsymbol{\Lambda}] \cdot \boldsymbol{\varphi}_{\alpha,k}(t) , \qquad (3)$$

where $\Delta_\alpha(t)$ is the bias potential of the lead $\alpha$; $\boldsymbol{\Lambda}_\alpha$ is the linewidth function of the lead $\alpha$ with the definition $\boldsymbol{\Lambda}_\alpha = -\text{Im}[\boldsymbol{\Sigma}_\alpha^r(\varepsilon=0)]$ and $\boldsymbol{\Lambda} = \sum_\alpha \boldsymbol{\Lambda}_\alpha$; $\mathbf{I}$ is a unit matrix; $z_k$ and $R_k$ are the $k^{\text{th}}$ Padé pole and coefficient for the Padé expansion of the Fermi-Dirac function [21, 23]

$$f_\alpha(\varepsilon) = \frac{1}{2} - \frac{1}{\beta}\sum_k^{N_k}[\frac{R_k}{\varepsilon - \Delta_\alpha - z_k/\beta} + \frac{R_k}{\varepsilon - \Delta_\alpha + z_k/\beta}] \qquad (4)$$

with $\beta$ being the inverse temperature. We see that $\boldsymbol{\Lambda}_\alpha$ is related to the imaginary part of self-energy matrix at the zero energy, which is a constant for the WBL approximation. Eqs. (1)-(3) are the central equations for the HEOM-WBL theory.

The initial values of $\boldsymbol{\sigma}(t)$ and $\boldsymbol{\varphi}_{\alpha,k}(t)$ are obtained from the residue theorem with the following expressions [21]

$$\boldsymbol{\sigma}(0) = \frac{1}{2}\mathbf{I} + \sum_{k=1}^{N_k}\text{Re}\{(\frac{-2R_k}{\beta})[(z_k/\beta)\mathbf{I} - \mathbf{h}_D(0) + i\boldsymbol{\Lambda}^z]^{-1}\} , \qquad (5)$$

$$\boldsymbol{\varphi}_{\alpha,k}(0) = \frac{2iR_k}{\beta}[(\Delta_\alpha(0) + z_k/\beta)\mathbf{I} - \mathbf{h}_D(0) + i\boldsymbol{\Lambda}^z]^{-1} \cdot \boldsymbol{\Lambda}_\alpha^z . \qquad (6)$$

After calculating the initial state of the system, some iteration methods (such as 4$^{\text{th}}$-order Runge-Kutta scheme) are to be used to investigate the dynamic process of these quantum open system.

### C. The Coulomb interaction

In our systems, electron is excited by the light to the conduction band and hole forms in the valence band. In the real space, the fluctuated charges on each atom site induced by the radiation also have the additional Coulomb interactions with each other. We consider the Coulomb interaction by the following formula [26-28]

$$E = \sum_i^{occ} <\Psi_i | H_0 | \Psi_i> + \frac{1}{2}\sum_{\alpha,\beta}^N \gamma_{\alpha\beta}\Delta q_\alpha \Delta q_\beta , \qquad (7)$$

where $\Delta q_\alpha = q_\alpha - q_\alpha^0$, is the Mullikan charge difference between the charged and neutral atom $\alpha$; $\gamma_{\alpha\beta}$ is the quantity to measure the electron-electron interaction. $\gamma_{\alpha\beta}$ is a function of the distance between atom $\alpha$ and $\beta$, which decays as $1/|\mathbf{R}_\alpha - \mathbf{R}_\beta|$ at large distances between two atoms and behaves as the Hubbard-like term $\gamma_{\alpha\alpha} = U_\alpha$ when $\mathbf{R}_\alpha = \mathbf{R}_\beta$ in the on-site case. The Hubbard parameter $U_\alpha$ is fitted from the DFT calculation for the spin unpolarized electron systems[27].

## III. RESULTS AND DISSCUSSIONS

In our calculations, the electron-photon interaction (EPI) Hamiltonian is utilized as[29]

$$H_{i,j}^0 e^{i\int_{\mathbf{R}_i}^{\mathbf{R}_j} \frac{e}{\hbar}\mathbf{A}d\mathbf{r}} = H_{i,j}^0 e^{i\frac{e}{\hbar}\mathbf{A}(\mathbf{R}_i - \mathbf{R}_j)}$$

where $\mathbf{A}$ is the vector potential along the molecular wire direction. We see that the electromagnetic (EM) field causes an additional phase factor on the Hamiltonian. This exponent factor is called the Peierls phase factor $\varphi_P$. For the PA chain, only the nearest-neighbor hopping terms are non-zero, and we assume $|\mathbf{R}_i - \mathbf{R}_{i\pm 1}| = a$ in the TB model, where $a$ is the average distance of the carbon chain. So we have $\varphi_P = \frac{e}{\hbar}Aa = \varphi_0 \cos(\omega t)$. With the Coulomb gauge, $\mathbf{A}$ is related to the electric field $\mathbf{E}$: $\mathbf{A} = \mathbf{E}/\omega$. Although the electric field here is a non-conservative field, we still may define an equivalent electric potential (EEP) $V_m$, which is related to the electric field by: $\mathbf{E} = V_m/(Na)\mathbf{e}_E$, where $N$ is the number of atoms along the molecular wire, $\mathbf{e}_E$ is the vector unit of $\mathbf{E}$. Then we have

$$\varphi_0 = \frac{e}{\hbar}(\frac{1}{\omega}\frac{aV_m}{Na}) = \frac{e}{\hbar}(\frac{aV_m}{Na})(\frac{\hbar}{e\omega_V}) = \frac{V_m}{N\omega_V}, \tag{8}$$

where $\omega = \omega_V \frac{e}{\hbar}$, $\omega_V$ is the dimensionless angular frequency.

### A. Electron and hole evolution in PA systems under a uniform radiation

Firstly, we calculate for the occupation number changes of the molecular orbitals in a finite PA chain under a uniform EM radiation. For such finite molecular chain, we obtain the eigenstates $\phi_n$ and the density matrix $\sigma$ from the eigenvalue calculation $\mathbf{H}_0\phi_n = E_n\phi_n$; $\sigma_{i,j} = \sum_{k}^{occ}\phi_{ik}^*\phi_{jk}$ [30]. Then we use the equation of motion for the evolution of the density matrix

$$i\dot{\sigma}(t) = [\mathbf{h}_D(t), \sigma(t)]. \tag{9}$$

Finally a matrix transformation $\sigma_M = \mathbf{F}^t\sigma\mathbf{F}$ is employed to obtain the orbital occupation for the electron. Here $\mathbf{F}$ is the transformation matrix between the atomic orbital basis and the molecular orbital basis, which means $F_{i,j} = \phi_{i,j}$, where $\phi_{i,j}$ is from the eigenvalue calculation.

Figures 2(a)-(d) show the calculated molecular orbital population from the evolved electron density matrix at different excitation energy of EM wave. If a continuum-wave (cw) laser has the power about 8.2W, and we assume the laser spot has the radius of $2\ \mu m$. According to the light intensity formula $I = \frac{1}{2}c\varepsilon_0 E^2$, we can evaluate the electric field is $2.2*10^7$ V/m, the EEP ($V_m$) along the PA chain is 0.12V. Now we calculate the EOM of electron density matrix by Eq. (9) with the Peierls factor for the EPI.

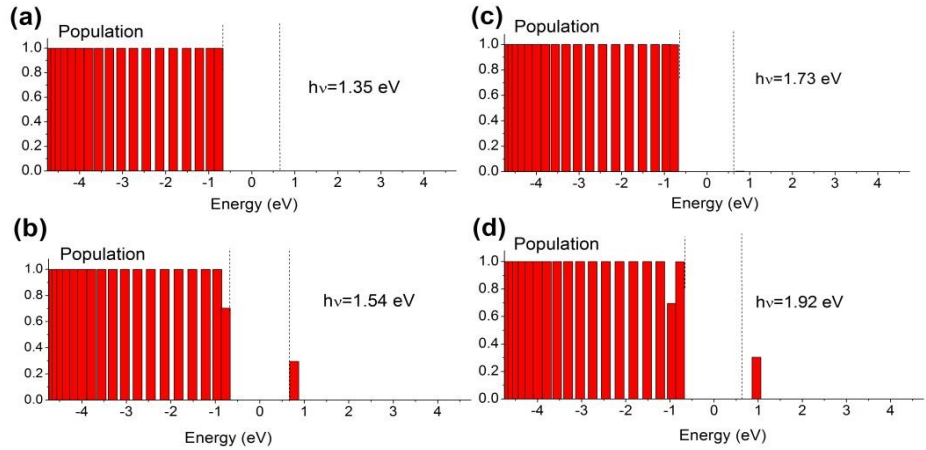

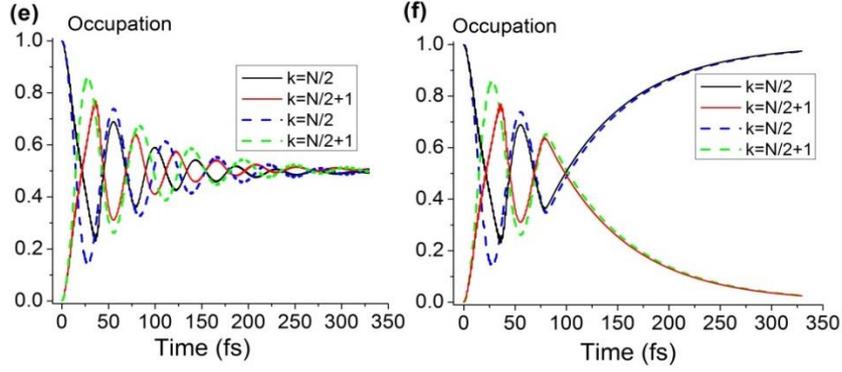

**FIG. 2.** The electron population on molecular orbitals for a short PA system (N=38) under a spatial-uniform EM radiation. The electric potential of the light is 0.1 V. (a)-(d) show the orbital distribution under different EM frequencies: (a) $\hbar\omega$ =1.35 eV; (b) $\hbar\omega$ =1.54 eV; (c) $\hbar\omega$ =1.73 eV; (d) $\hbar\omega$ =1.92 eV; The two dashed lines in the figures indicate the edges of the valence and conductance bands. (e) and (f) show the two molecular orbitals' evolution under a constant EM radiation (see (e)) and a finite-time EM radiation (see (f)). The two solid lines correspond to the system with the Coulomb interaction and the two dashed lines correspond to the system without the Coulomb interaction.

In Fig. 2 (a) we see that when the PA chain is illuminated by light with the energy smaller than the band-gap energy $E_g$ (1.54 eV), there is almost no electron transition. When the energy of the illumination light (1.54eV) approaches or become a little larger than $E_g$, there are apparent transitions (Fig. 2(b)). The occupation of the first HOMO orbital (with the energy $E_{-1}$=-0.77 eV) decreases to about 0.70; and the occupation of the first LUMO orbital (with the energy $E_1$=0.77eV) increase to about 0.30. In Fig. 2(c) when the illumination light has the energy of $E_2 - E_{-1}$ or $E_1 - E_{-2}$ (1.73 eV), there are no corresponding optical transition between these two levels. This is due to the selection rule for the dipole radiation [31]. In a PA system the orbitals have the similar forms as the eigenfunctions in a 1D infinite potential well. They have even or odd symmetry with the wavefunctions as $\phi_n(x) = sin(\frac{n\pi}{L}x)$. So the transition matrix $<\phi_n|x|\phi_m>$ gives the selection rule: only when *n-m* is equal to an odd number, can the transition occur. Consequently, the transitions such as $E_2 \leftrightarrow E_{-1}$ or $E_1 \leftrightarrow E_{-2}$ are forbidden. In Fig. 2(d) the energy of illumination light is 1.92 eV, which is about the energy difference between $E_{-2}$ and $E_2$. We see there is an apparent transition since it obeys the selection rule.

Furthermore, we use this basis transformation idea to investigate the orbital evolutions of an open system. The device is a PA chain with 38 carbon atoms sandwiched by two carbon chains (see Fig. 1(a)). We set a weak coupling between the device and the leads (with a hopping energy $t$=-0.7 eV). The molecular orbitals of the device (regarded as an isolate system due to the weak coupling) are calculated at first. Then the WBL-HEOM calculation is employed to for the system under an EM radiation and the evolved density matrix is transformed to the molecular basis. A constant radiation (see Fig. 2 (e)) and a finite-period radiation (see Fig. 2 (f), the radiation period ranges from 0 to 80 fs) are employed with the EM energy of 1.54 eV.

In Fig. 2(e), we see that the HOMO and LUMO orbitals oscillate and then tend to about 0.5. The oscillation amplitude decays with time like a damped Rabi model [32]. In this process the electrons transit mainly between HOMO and LUMO orbitals and exchange energies with the external EM wave. The decay is due to the dissipation source of two leads. This is different from the isolate system, in which the electrons will periodically transit between HOMO and LUMO orbitals as the Rabi cycle. In a long time limit the generation and dissipation of these two orbitals reaches an equilibrium state. In Fig. 2(f) we see that after the EM radiation is turned off, the HOMO orbital occupation quickly goes back to 1 and the LUMO orbitals occupation goes back to 0. This is because the oscillation charges tend to dissipate into the two leads due to the spread of the waves and in the end they return to the initial equilibrium state without radiation. We observe that in an isolate system all the occupations will become unchanged when turning off the radiation, because the inertial charge oscillations would not change the orbital occupations.

Here we also note that this dissipation is similar to the electron-hole recombination process: after the recombination, the electron in the conduction band (LUMO) returns to the valance band (HOMO). In general solar cells the recombination energy is taken away by phonons (the Schockley-Read-Hall process [9]); photons (radiative recombination due to the spontaneous emission[33]) or electrons (Auger recombination[34]). All these cases lead to the dissipation of the excited electrons and holes. In this PA system the dissipation is caused by the coupling of the leads, due to the spread nature of the electron waves.

At last, from Fig. 2(e) and (f) we see that when involving the Coulomb interaction, the population fluctuations (solid lines) becomes smaller than the non-Coulomb-interaction case (dashed lines). This is due to the electric attraction of opposite charges, which will be detailed later.

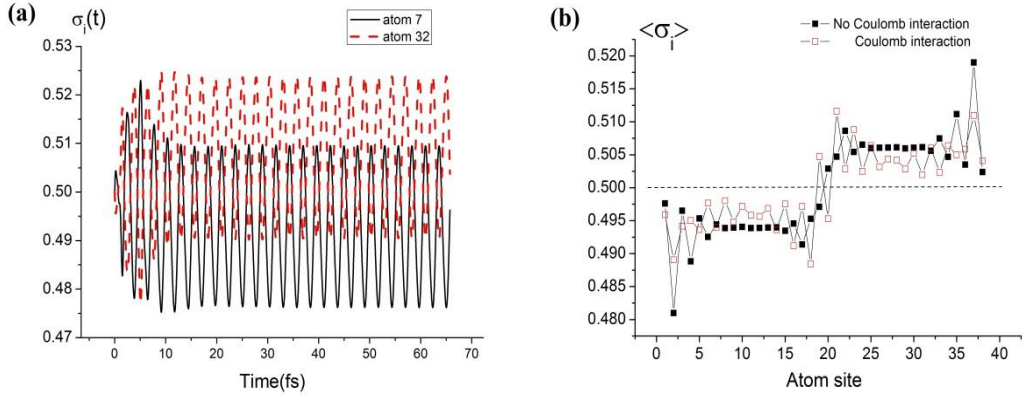

**FIG. 3.** The electron density situation of a PA chain under a uniform EM radiation with the radiation energy being $\hbar\omega$ =1.55 eV. (a) The electron density dependence on time in different atoms of a PA system (the black solid line is for atom 7 and the red dashed line is for atom 32); (b) The time-averaged electron density distribution on the PA chain. The filled-square line is without the Coulomb interaction; the empty-square line is with the Coulomb interaction. In (a) and (b), the PA has N=38 atoms. The two bias potentials ($\pm 0.2$ V) are applied on the two sides of PA chain.

Now we use our WBL-HEOM method to calculate for the charges of a PA chain in real space. The PA chain, sandwiched by two carbon leads, is doped with n and p types which is the same as the case in photovoltaic devices. To obtain a large charge fluctuation, we use the pulse-typed laser with a high electric field. The pulse with a duration time of 150 fs and the repeating frequency of 80 MHz is applied from the Ti-sapphire laser in the experiment. The laser energy is about 0.1 nJ per pulse and the spot radius is 2 $\mu m$. The electric field can be evaluated as 2.0*10$^8$ V/m and the EEP ($V_m$) along the PA chain is 1.0 V. In the WBL-HEOM simulation, EEP rises smoothly from the zero initial value to a sinusoidal function with the amplitude $V_m(t) = V_m(1 - e^{-t/\tau})\cos(\omega t)$.

Figure 3 (a) shows several electron density variations versus time under a uniform EM radiation. The PA system has two bias potentials on the left and right regions (see Fig. 1(c)). We see the two atoms on the left side (atom 7) oscillate with an average value below 0.5; and the two atoms on the right side (atom 31) oscillate with an average value above 0.5. This is because that the negative bias on the left atoms leads to less electrons on them, while the positive bias on the right side attracts more electrons. Figure 3(b) gives the average electron density distribution in the PA chain under the EM radiation with the energy of 1.55 eV (filled-square line) without considering the Coulomb interaction. For the same reason, we see the atoms in the left side almost have the lower electron densities than the atoms in the right side. If the Coulomb interaction is considered, the density (the empty-square line) deflection from the neutral density (0.5) becomes small due to the attraction between positive and negative fluctuation charges.

### B. Electron-hole separation in PA chains

Then we use a longer PA chain (N=160) to study the dynamics of electron-hole separation. The EM pulse is radiated on the center region of the PA chain. As shown in Fig. 1(d)-(f), the radiation electric field is a Gaussian-typed pulse both in the time and spatial domain, $V_{ac}(x,t) = V_0(x) e^{-(t-t_0)^2/\tau^2} \cos(\omega t)$, where the spatial dependence of the electric field $V_0(x)$ is given as $V_0(x) = V_m e^{-(x-x_0)^2/A^2}$ for a pulse or $V_0(x) = V_m$ for a uniform EM wave. In our simulation, we set $t_0$ =6 fs; $\tau$ =2.5 fs; $x_0 = Na/2$. The energy of the EM wave is 2.5 eV. The laser energy is 7.8 nJ per pulse with 150 fs duration time and 80 MHz repeating frequency. The corresponding average power is 0.63 W and the equivalent electric potential along the PA chain is 40 V.

Figures 4 (a)-(d) and (e)-(h) show four snaps of the averaged charge distribution at the time 4.0 fs, 8.0 fs, 13.2 fs and 18.5 fs. The PA chain has the uniform bias potential as shown in Fig. 1(d). The bias is set as $\Delta = \pm 0.4$ V. In Figs. 4 (a)-(d), the Coulomb interaction is not considered while in Figs. 4 (e)-(h) the Coulomb interaction is involved. Due to the frequent charge fluctuations in time and space domain, some averaging implementation can be done to obtain the effective charge distribution. This average processing in a small period (about one period of the EM wave) eliminates the transient fluctuation effect and shows the intrinsic charge properties (such as Fig. 3).

If the averaged value is negative, the atom is electron-like at that moment; and if the averaged value is positive, it corresponds to a hole-like atom.

We see at the beginning the charge oscillations rise in the middle part of the PA chain. Then the oscillations tend to form the negative and positive fluctuation charge on the left and right side respectively. These two types of electron wave packet later move towards two sides with certain spread. At last these wave packets meet and dissipate into the two leads and disappear from the device region. The same profile of wave packets for the electron and hole is due to the symmetry of the conduction and the valence bands in the tight-binding model. It is noticed that the Coulomb interaction makes the charge fluctuation much smaller but more frequent in space, and the wave packets in Figs. 2 (e)-(h) have a wider spread. This wide spread makes a slower separation speed compared to the case without the Coulomb interaction.

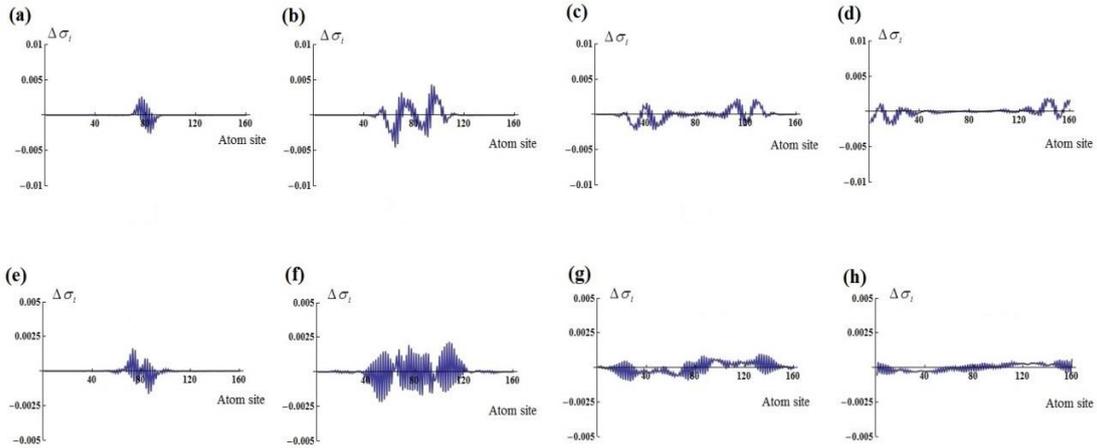

**FIG. 4.** Snaps of the electron and hole evolution in the PA system (N=160, with the averaged continuous impurity potential: $\Delta = \pm 0.4$ eV) under a EM pulse ($\hbar\omega$ =2.5 eV) radiated in the middle region of the system. In the upper panel no Coulomb interaction is considered; in the lower panel Coulomb interaction is considered. From (a)-(d) in the upper panel or (e)-(h) in the lower panel, the snap is taken at time=4.0 fs; 8.0 fs; 13.2 fs; 18.5 fs respectively.

Similarly, we also use the WBL-HEOM scheme to calculate for the charge separation process for a PA chain with two types of doping atoms. The doping atoms have positive or negative on-site potentials ($\Delta = \pm 2.0$ V) and are separated by the same number of carbon atoms on the two sides (see Fig. 1(e)). The EM radiation and other parameters are the same as in the previous case. Fig. 5

(a) shows two snaps of fluctuation charge with the doping atoms in every 4 carbon atoms. Fig. 5 (b) shows the similar snaps with the doping in every 10 atoms. We see that the electron-hole wave packets in these two cases almost have the similar profiles and the charge fluctuation in the low doping concentration (every 10 atoms) is smaller.

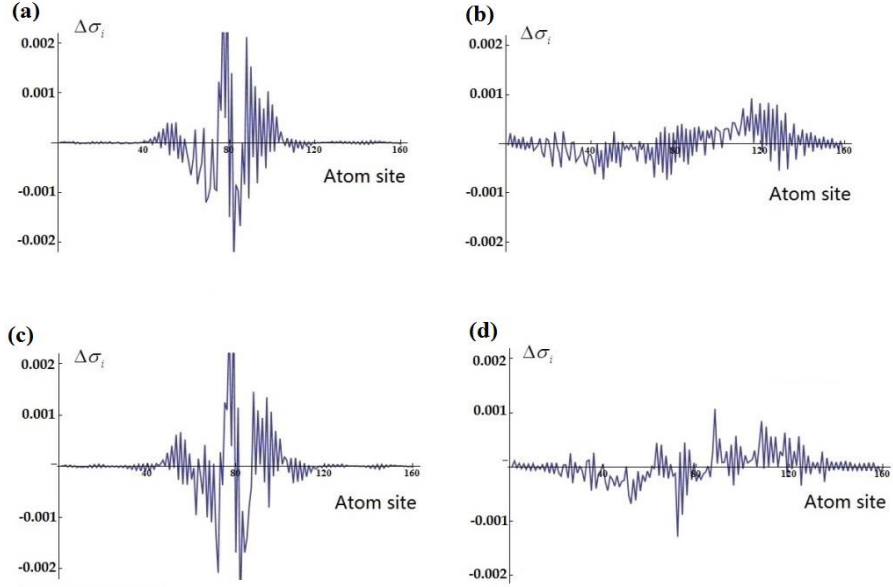

**FIG. 5.** Snaps of the electron and hole evolution in the PA system under a EM pulse ($\hbar\omega$ =2.5 eV) radiated in the middle region of the system. The snap time is 6.6 fs ((a),(c)) and 14.6 fs ((b),(d)). In (a) and (b), the impurity atoms locate in every 4 sites with the potentials: $\Delta = \pm 2.0\,\text{eV}$ in the left and right parts; In (c) and (d) the impurity atoms locate in every 10 sites with the potentials: $\Delta = \pm 2.0\,\text{eV}$ in the left and right parts.

Now we investigate the electron-hole separation speed under different illumination intensities. When increasing the light intensity (or the excitation electric potential), the charge fluctuation becomes larger. Due to the Coulomb attraction between the electrons and holes, we expect that the wave packets under a strong light intensity will spread slower than those in a weak light condition. However, we find that the previous average processing scheme is not suitable here. This is because the fluctuation effect is eliminated by averaging in a time period. This effect just reflects wave packet speed. To observe the wave packet evolution clearly, we draw the envelope lines of the wave packet by taking out the maximum amplitude of charge on each atom during one EM period.

Figure 6 shows such results for the weak and strong light cases. It is easy to see that under a strong light illumination (Figs. 6(e)-(h)), the large charge fluctuation occurs and the Coulomb interaction make the wave packet a slower spread behavior, compared to the envelope lines under a weak light illumination (Figs. 6(a)-(d)).

This is similar to the classical simulation and the experimental measurements for silicon nanowires [7]. In the semi-classical drift-diffusion model, the electrons and holes move driven by the electric field (drift part) and the concentration gradient (diffusion part). The electric field is determined from the Poisson equation which is determined by the total charges and external voltages. Hence, the large fluctuated charge leads to a large Coulomb attraction, which makes a slow separation for the two opposite charge distributions in the nanowire [7]. Our calculation is about the ballistic motion for the coherent electrons. In this quantum regime, the kinetic motion of the wave packets plays a major role. Only when the Coulomb interaction energies are comparable to the hopping (kinetic) energies, the Coulomb-related charge separation reduction can occur. This explains the very large charge fluctuations (about 0.1-0.2) in Fig. 6 (e)-(h) for the electron-hole localization under a strong light.

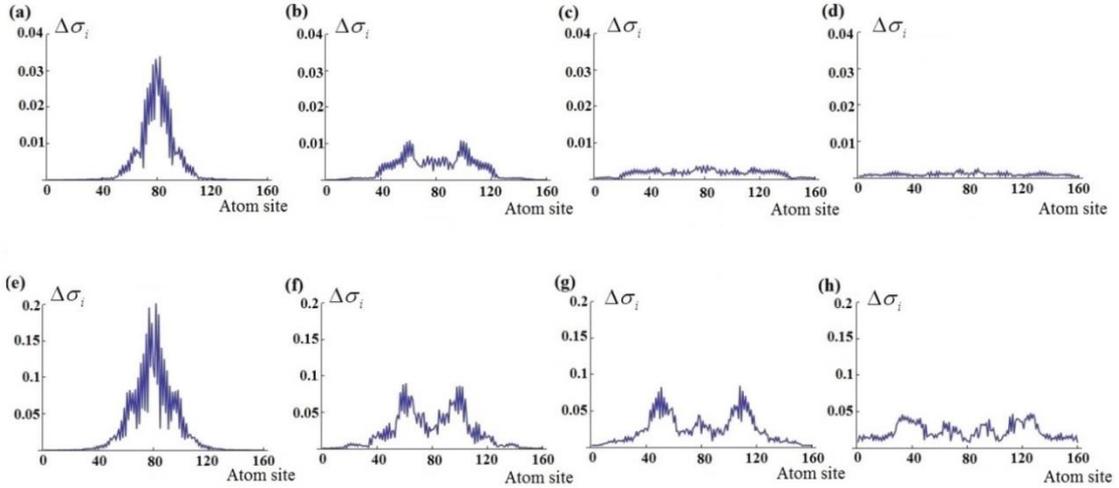

**FIG. 6.** The charge spread dynamics under a weak laser light (7.8 nJ per pulse or EEP=40 V) and strong laser light (12.5 μJ per pulse or EEP=1600 V) in a p-i-n typed PA chain (N=160, with the averaged impurity potentials: $\Delta = \pm 0.4$ eV). In the upper panel the snaps of charge evolution at 5.3 fs (a); 7.9 fs (b); 10.5 fs (c) and 13.2 fs (d) are shown. In the lower panel, the snaps of charge evolution at 5.3 fs (e); 7.9 fs (f); 10.5 fs (g) and 13.2 fs (h) are shown.

## C. Eectron-hole separation in PP and doped PP systems

In the last section, we apply this WBL-HEOM method to PP systems for investigating the charge separation. The PP chain consists of 60 benzene units as shown in Fig. 1(b). A bias potential of +/-0.4 V is applied on the two sides to stand for the n-dope and p-dope regions. For the laser, the energy ($\hbar\omega$) is 2.7 eV and EEP ($V_m$) is 8.0 V. Other parameters about the pulse are as same as before. Similar to Fig. 4 and Fig. 5, a short-time-average scheme is used to show the electron-hole evolutions in the PP chain. In Fig. 7 we see an apparent positive and negative charge separation and the movement towards the two sides after a pulse is radiated in the middle of the chain. After some time about 20 fs, the separated charges reach the lead regions and dissipated out. The involution of the Coulomb interaction also makes the spread of the two wave packets.

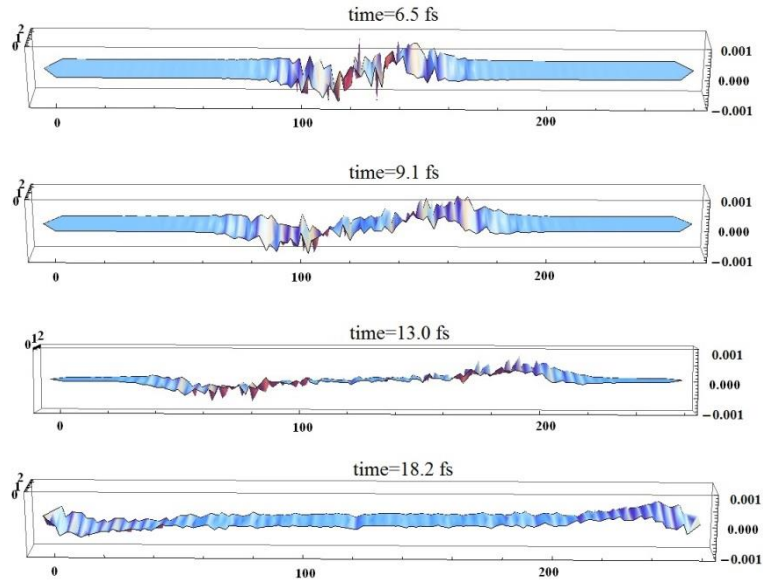

**FIG. 7.** Snaps of the charge separation process in the PP chain (60 benzene units) under a laser pulse illuminated in the middle. The bias potentials on the two sides of PP chain is +/- 0.2 V and the EEP of the EM radiation is 8.0 V.

Finally, we choose another PP chain system doped with boron and nitrogen atoms as described in Fig. 1(c). We design a heterocyclic ring as the unit of a molecular wire. In each unit three heterogeneous atoms (boron and nitrogen) are doped in one benzene ring and the first-principles calculations (DFTB method) are employed for the band structure. DFTB is an approximation of density-functional-theory (DFT) model derived from the second-order expansion of Kohn-Sham energy around the reference charge density [26-28]. A free software package: DFTB+ [26, 27] developed by the Th. Frauenheim's group is used in our paper. The vacuum layers of 3.9 nm in the ribbon plane and 50 nm in the normal direction is applied to avoid the interaction between ribbons in different unit cells. In the SCC iteration we choose the Broyden mixing scheme and the energy tolerance is set to $1*10^{-6}$ eV. The geometric structure of the unit cell is optimized by the conjugate gradient algorithm with the Brillouin zone sampling of 100 points in x direction.

Figure 8(a) shows the DFTB band structure. We see that there exist an energy gap of 2.2 eV and the valence band curvature at k=0 is found to be larger than the conduction band curvature (with a ratio of 1.1). This means that the effective mass of hole in the valence band is smaller than that of electron in the conduction band. To calculate the dynamic process with our WBL-HEOM method, TB model is needed. With some suitable parameters similar to the previous study [35, 36], we obtain the band structure by the TB model (Fig. 8(b)). In this simple TB model we only change the on-site energies of carbon/boron/nitrogen atoms (we change the on-site energies of B and N atoms as $\varepsilon_B=\varepsilon_3=3.70$ eV; $\varepsilon_N=\varepsilon_1=\varepsilon_6=-0.85$ eV; the on-site energies of other C atoms are evaluated from all the contributions of the neighboring B/N atoms with decaying trends as shown in Fig.3 of reference [35]: $\varepsilon_2=0.90$ eV; $\varepsilon_4=1.46$ eV; $\varepsilon_5=1.56$ eV). Both bands are shifted to zero as its Fermi level. We see there exist differences between these two bands. However the basic band properties near the Fermi level remain the same: the TB band has a gap of 2.0 eV and the curvature of the valence band is larger than that of the conduction band (with a ratio of 1.2). Although the band gap and curvature ratio are a little larger than the DFTB values, we believe this TB model may effectively mimic the basic optical excitations for this heterocyclic system.

Similar to the setup in Fig. 7, we apply the uniform potentials (+/-0.2 eV) on the left and right parts of the molecular chain for mimicking the p-type and n-type doping. Then a laser pulse with an energy of 2.1 eV and EEP of 16 V is radiated on the middle of the system. The snaps with some time average are shown in Fig. 8(c). It is apparent to see that the negative wave fluctuation (corresponding to excited holes in valence band) has a much quicker separation than the positive wave fluctuation (corresponding to excited electrons in conduction band). This is expected as a result of a larger effect mass for the electron. And the hole-type wave is found to have a much larger spread behavior in this dynamic process. There are similar observations in the previous experiment in silicon nanowire [7, 8].

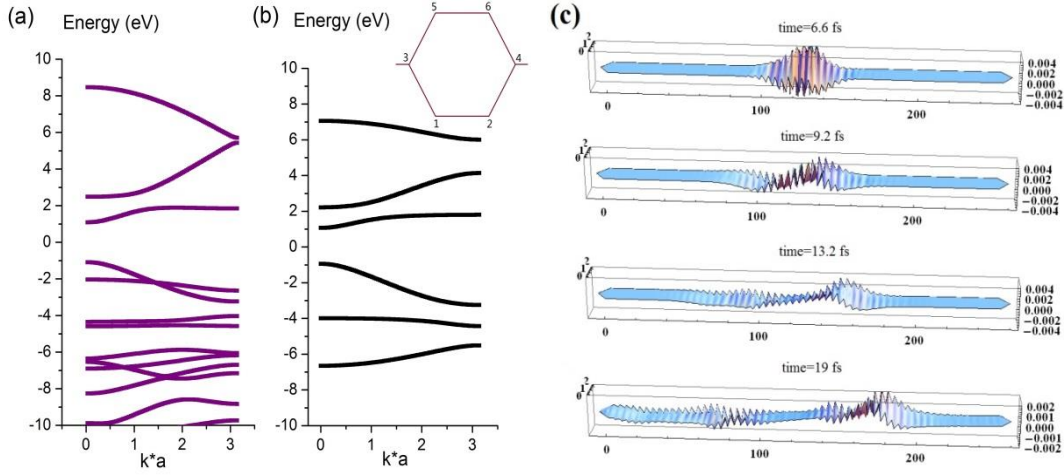

**FIG. 8.** DFTB (a) and TB (b) band strucutures of the boron-and-nitrogen-doped PP system (60 heterocylic ring units). The detailed atomic structure of the doped PP is shown in Fig.1(c). The inset figure shows the atom indexes of a heterocylic ring in the TB model. (c) Dynamic electron and hole separation processes in this doped PP chain under a laser pulse with the enregy of 2.1 eV and EEP of 16 V.

## IV.    CONCLUSION

The time-dependent quantum transport method is employed to molecular wire systems for investigating the dynamic charge dissipation and separation behaviors. The method is based on the previous developed HEOM theory with the wide-band-limit approximation. The Coulomb interaction is considered by the spin-unpolarized Hubbard model.

We calculate the molecular orbital occupation changes of a finite PA chain under a uniform EM radiation. Only the transitions obeying the selection rules can occur. For a weak-coupling PA system we observe a damped oscillation for the electron occupations of the valence and conduction bands. After the radiation is turned off, the electrons in the conduction band smoothly return to valence band due to the lead dissipations.

The dynamic wavepacket evolution is calculated in the PA and PP chain with some doping atoms under a laser pulse in the middle region. In PA chain, we find the electron and hole wavepackets has the same separation profiles with the averaging implementation. Involving of the Coulomb interaction leads to a small and frequent fluctuations and wide spread. In a strong illumination, the wave packet has a slower separation with the enveloping implementation. In PP systems, we find similar charge separation behaviors. For the doped PP chain, we find the hole wavepacket moves faster than the electron wavepacket due to its smaller effect mass.

So for these molecular wires, we employ the uniform EM radiation to investigate the dissipation dynamics in the mode space and we employ the pulsed radiation for the dissipation dynamics in the real space. We believe this HEOM-WBL method is a good tool to study the quantum wavepacket dissipation and separation in the open nano devices. Some potential factors such as electron-phonon interaction and radiative recombination may be involved in the future researches.

## ACKNOWLEGEMENTS

The authors are grateful to Prof. Guanhua Chen in the University of Hong Kong for useful discussions and help. We thank Dr. Yong Wang in Nankai University for the helpful discussions on the exciton theory, and Dr. Rui Wang in the Chongqing University for the kind help of the computer service. Financial support from the starting foundation of Chongqing University (Grants No. 0233001104429) is also gratefully acknowledged.

*Email: xiehangphy@cqu.edu.cn